\documentclass[sn-mathphys-num]{sn-jnl}

\usepackage{booktabs}\let\cline\cmidrule
\usepackage{graphicx}%
\usepackage{multirow}%
\usepackage{float}
\usepackage{amsmath,amssymb,amsfonts}%
\usepackage{amsthm}%
\usepackage{mathrsfs}%
\usepackage[title]{appendix}%
\usepackage{xcolor}%
\usepackage{textcomp}%
\usepackage{manyfoot}%
\usepackage{booktabs}%
\usepackage{algorithm}%
\usepackage{algorithmicx}%
\usepackage{algpseudocode}%
\usepackage{listings}%
\usepackage{adjustbox}
\usepackage{graphicx}

\theoremstyle{thmstyleone}%
\theoremstyle{thmstyletwo}%

\theoremstyle{thmstylethree}%

\begin{document}

\title[Cities Reconceptualized: Unveiling Hidden Uniform Urban Shape through Commute Flow Modeling in Major US Cities]{Cities Reconceptualized: Unveiling Hidden Uniform Urban Shape through Commute Flow Modeling in Major US Cities}


\author*[1]{\fnm{Margarita} \sur{Mishina}}\email{mishina@math.muni.cz}

\author[]{\fnm{Mingyi} \sur{He}}\email{mh5172@nyu.edu}

\author[2]{\fnm{Venu} \sur{Garikapati}}\email{Venu.Garikapati@nrel.gov}
\author[1,3]{\fnm{Stanislav} \sur{Sobolevsky}}\email{sobolevsky@nyu.edu}

\affil*[1]{Department of Mathematics and Statistics, Faculty of Science, Masaryk University, Brno, 611 37, Czech Republic}

\affil[2]{National Renewable Energy Laboratory. 15013 Denver West Parkway, Golden, Colorado 80401}

\affil[3]{Center for Urban Science and Progress, New York University, New York, NY, 11201, USA}


\abstract{Urban development is shaped by historical, geographical, and economic factors, presenting challenges for planners in understanding urban form. This study models commute flows across multiple U.S. cities, uncovering consistent patterns in urban population distributions and commuting behaviors. By embedding urban locations to reflect mobility networks, we observe that population distributions across redefined urban spaces tend to approximate log-normal distributions, in contrast to the often irregular distributions found in geographical space. This divergence suggests that natural and historical constraints shape spatial population patterns, while, under ideal conditions, urban organization may naturally align with log-normal distribution. A theoretical model using preferential attachment and random walks supports the emergence of this distribution in urban settings. These findings reveal a fundamental organizing principle in urban systems that, while not always visible geographically, consistently governs population flows and distributions. This insight into the underlying urban structure can inform planners seeking to design efficient, resilient cities.}




\maketitle

\section{Introduction}\label{sec1}

As urbanization continues to accelerate worldwide, cities are becoming increasingly important centers of economic activity, cultural exchange, and social interaction. This rapid transformation emphasizes the need to study the patterns and principles that drive urban growth and population movements. Having a deep understanding of these factors becomes crucial in tackling the global challenges of sustainable urban planning and infrastructure development in both established and emerging regions. 

Over the decades, researchers have sought to uncover the fundamental principles underlying urbanization trends. Through their efforts, they have identified striking similarities in the ways populations move and settle within urban spaces. One of the earliest discussions in this area focused on the statistical distribution of populations and economic activities across cities. In the early twentieth century, it was observed that city sizes followed a power-law distribution \cite{auerbach1913gesetz}, a relationship later formalized by Zipf's Law \cite{gabaix1999zipf}, which suggests that a city's size is inversely proportional to its rank. However, an alternative perspective soon emerged, proposing that city sizes are better modeled by a log-normal distribution, suggesting that urban growth results from the multiplicative effects of many random and independent processes \cite{levy2009gibrat}. The comparison between these two models has been the subject of numerous studies \cite{malevergne2009gibrat, gonzalez2015size}. While Zipf's Law has proven surprisingly effective in revealing regularities among cities with populations above a certain threshold \cite{gabaix2004evolution}, the log-normal distribution has been shown to provide a better fit for smaller cities and city units \cite{decker2007global} or when considering cities within a certain geographic proximity \cite{gonzalez2019lognormal}. Further research has leveraged the strengths of both models by proposing a hybrid approach that combines a power-law distribution for large cities with a log-normal distribution for smaller ones. 

While much attention has been devoted to understanding population distributions at the city level, smaller urban units --- such as neighborhoods, counties, and administrative divisions --- have received less focus. This gap can be attributed to two primary challenges: a historical lack of reliable data at such granular levels, and the inherent complexity of cities themselves. Although recent advances in data collection now allow for more detailed analysis, cities are still shaped by a unique blend of historical, geographical, and economic factors. This complexity makes it difficult to generalize findings across different contexts, often leaving urban planners to adapt strategies for each city's distinct characteristics.

In addition to city size distributions, researchers have actively explored patterns of population mobility, developing spatial interaction models to predict human movement and settlement. Two major schools of thought have emerged in this area. The first was inspired by Zipf's adaptation of Newton's law of gravitation \cite{zipf1946p} and suggested that the number of trips between two locations is proportional to their population sizes and inversely proportional to the distance between them. The second was based on Stouffer's theory of intervening opportunities and proposed that the likelihood of travel between two locations is inversely related to the number of alternative destinations in between. Both models have gained widespread acceptance due to their simplicity and applicability across various contexts. However, with the growing quest for accurate prediction, recent advances in deep learning have led to the development of more sophisticated mobility models that capture the complex, non-linear patterns of human travel behavior.

These technological advancements have opened new avenues for studies that leverage the mobility models to gain deeper insights into the organization and growth of cities through urban dynamics. Recent studies \cite{wu2022multi, mai2022review} employ such models to learn the embeddings of urban locations from human mobility data to further use them in various downstream tasks. Specifically, in one study \cite{yao2018representing}, the authors developed a method that learns the representation of city zones by exploiting large-scale taxi trajectories to identify functional regions in New York. Another study \cite{fan2024neural} performed cauterization of location embeddings derived from population mobility data to reveal similarities among urban areas in order to support more targeted and effective urban planning and policy-making. The central idea that runs through much of similar research suggests that while identifying commonalities in urban layouts may be difficult through traditional geographic analysis, such patterns become clearer when examined through multidimensional embedding spaces.

Building on this concept, our research seeks to fill the gap in existing studies on universal principles that govern spatial city organization. Specifically, in the paper, we leverage a gravitational mobility model to learn a new two-dimensional representation of census tracts in major US cities, based on observed spatial population distributions and intra-city commuting flows. We find that in this redefined space adjusted to the mobility commute network, the spatial population distribution follows an ideal log-normal shape, a pattern that is not evident in traditional geographic context. Along with this empirical finding, we offer a theoretical explanation for the emergence of log-normal law in spatial population distribution, based on a simulation with preferential attachment random-walk model of urban settlement. 

The implications of our approach extend beyond merely identifying statistical patterns --- it reveals unifying principles of urban organization that transcend the natural limitations of geospatial layouts. By moving beyond the constraints of physical geography, this framework highlights deeper structural regularities in how populations are distributed and move within cities. Such insights are crucial for urban planners and policymakers, as they provide a new lens through which to understand urban growth and optimize the design of infrastructure and services. For instance, by using this model, urban planners can identify functional zones within cities that may not be apparent through traditional analysis, allowing for more efficient transportation networks, zoning laws, and resource allocation. Ultimately, this approach offers a practical tool for enhancing urban planning and development, fostering more sustainable and well-organized cities that can better accommodate future growth.

\section{Results and Discussion}\label{sec2}

\begin{figure}[H]
\centering
  \includegraphics[width=1\textwidth]{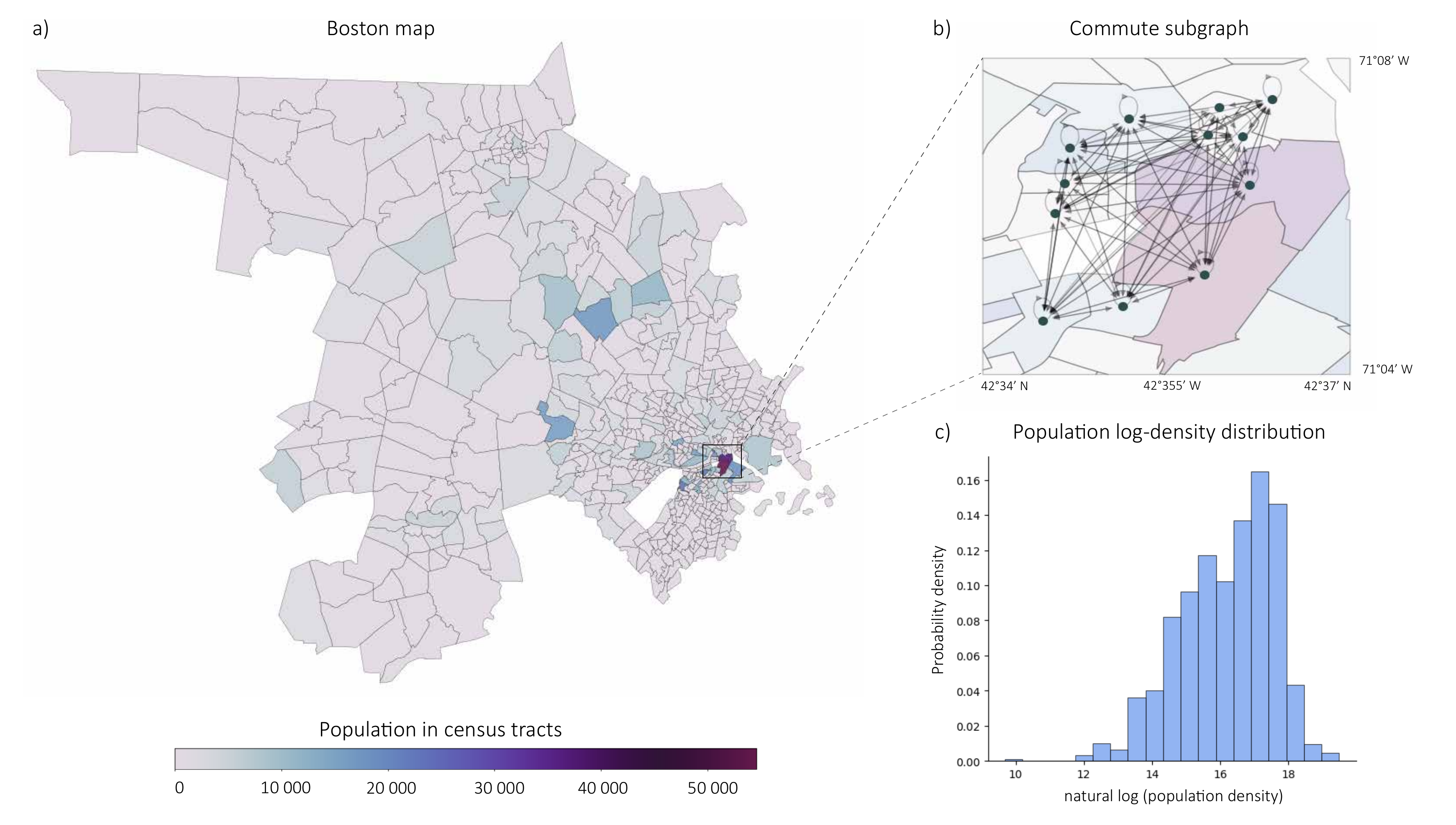}
\caption{Population log-density distribution with fitted normal curve.  \textbf{a.} Map of census tracts in Boston. \textbf{b.} Commute subgraph covering census tracts in the center of the city. \textbf{c.} Population log-density distribution in Boston.}
\label{fig:1_} 
\end{figure}

In this study, we analyzed data from 12 major US cities using the Longitudinal Employer-Household Dynamics (LEHD) dataset, which provides detailed employment, residence, and commute information for US workers, as published by the US Census Bureau \cite{LEHDData}. For each city, we constructed a commute network to represent the flow of workers between their places of residence and employment. In these networks, each node corresponded to a census tract, approximated by its centroid and characterized by its geographical area and population count. The connections between nodes, represented by weighted edges, indicated the volume of daily commuters traveling between home and work. The network also included loops, signifying cases in which both the place of employment and residence were located within the same census tract~(Fig.~\ref{fig:1_}).

For each census tract, we calculated the population density by dividing the population count by the census tract's area. In the original geographical space, the population density within almost all cities significantly deviated from the logarithmic normal distribution (Eq. \ref{eq:log_normal_pdf}), which typically characterizes many naturally occurring phenomena, including income levels and city sizes. This deviation suggests that the underlying social, economic, or infrastructural factors inherent to each city might influence the population distribution in a way that leads to more complex or skewed patterns than the random multiplicative processes that would result in an ideal log-normal distribution. Specifically, this can be rooted in established zoning laws, historical development patterns, characteristics of transportation networks, existing economic disparities, and other factors that concentrate populations in certain regions. The only exception to this observation was Chicago which population density distribution across census tracts initially followed a log-normal distribution. This anomaly can be attributed to Chicago's relatively late and organized development, flat geography, central industrialization, and more uniform infrastructure, which allowed for a more even spread of population across the city's census tracts compared to other cities.

\begin{equation}
   f(x; \mu, \sigma) = \frac{1}{x \sigma \sqrt{2 \pi}} \exp \left( -\frac{(\ln x - \mu)^2}{2 \sigma^2} \right), \quad x > 0
   \label{eq:log_normal_pdf}
\end{equation} 
where $x$ --- is population density in a census tract, $\mu$ and $\sigma$ --- are the mean and the standard deviation of the natural logarithm of the population density.
\\

To reduce the impact of historical and geographical constraints in the other US cities, we projected the census tracts into a new embedding space that was designed to better reflect commuting patterns across each city.  To this end, we employed the Origin-Constrained Gravity Model, a gravitational spatial interaction model that predicts population flows between two locations based on the attractiveness of the destination and the distance between them. The attractiveness of each destination was estimated based on the number of job opportunities, quantified as the number of people working there. The influence of distance was modeled using an exponential decay function with adjustable parameters $\alpha$ and $\beta$, which control how the likelihood of commuting decreases with distance. A distinguishing feature of this origin-constrained model is that it ensures the total number of commuters leaving each residential area matches the known number of residents commuting to work from that area.

\begin{equation}
    T_{ij} = O_i \cdot \frac{D_j \exp(\alpha d_{ij}^\beta)}{\sum_{k} D_k  \exp(\alpha d_{ij}^\beta)}
\end{equation}

where $T_{ij}$ --- the number of commuters from census tract~$i$~(living area) to census tract~$j$~(work area). $O_i$ --- the total number of residents in census tract $i$ who commute to work. $D_j$ --- the number of people working in census tract $j$. $d_{ij}$ --- the Euclidian distance between centroids of census tracts $i$ and $j$.
\\

In the first step, we fitted the model to the observed data on mobility flows by learning the parameters $\alpha$ and $\beta$ through the back-propagation of the gradients of the binomial likelihood function. After that, we allowed the model to fine-tune the coordinates of location centroids in a way that would better reflect commuting behaviors rather than actual geographical boundaries.

\begin{figure}[H]
\centering
  \includegraphics[width=1\textwidth]{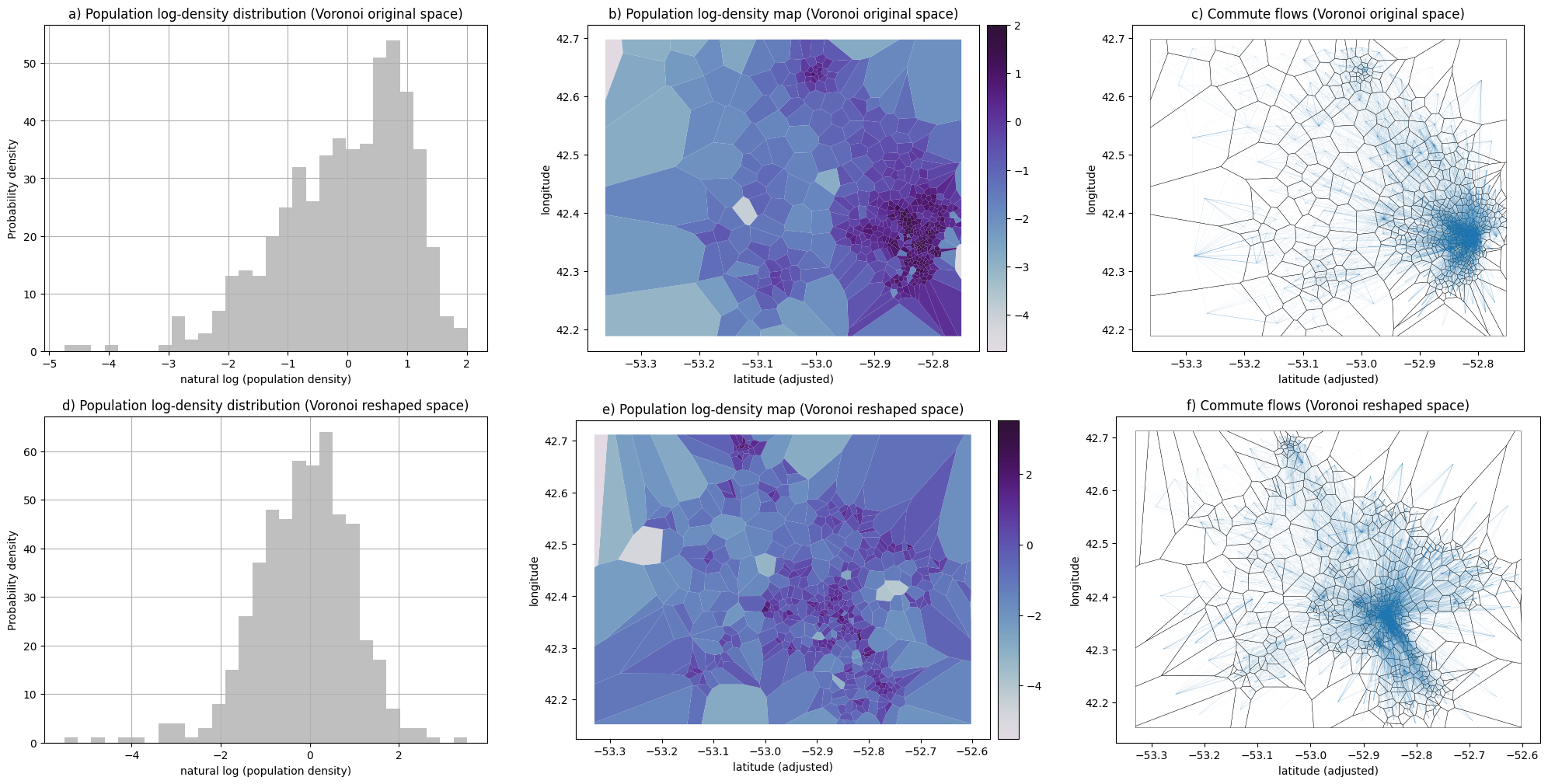}
\caption{Urban Density and Commute Flow Patterns in Geographic vs. Embedding Space.  \textbf{a.} Population log-density distribution across Voronoi regions in the original geographic. \textbf{b.} Map of the population log-density distribution across Voronoi regions in the original geographic layout. \textbf{c.} Commute flow network overlaying the original Voronoi tessellation. \textbf{d.} Population log-density distribution across Voronoi regions in the reshaped space. \textbf{e.} Map of the population log-density distribution across Voronoi regions in the reshaped space. \textbf{f.} Commute flow network overlaying the Voronoi tessellation in the reshaped space.}
\label{fig:2_} 
\end{figure}

Taking the learned embeddings as new location coordinates, we compared the distribution of population density over the original and new projected space.  To ensure consistent comparison of population densities in this reshaped space --- where geographical boundaries were altered --- we built Voronoi tessellations \cite{voronoi1908nouvelles} which divide a plane into regions based on proximity to a set of points. This method allowed us to uniformly calculate population densities, despite the altered spatial relationships in the embedding space. Once the geographical boundaries of urban locations were redefined, we observed that population densities in studied cities conformed closely to a log-normal distribution (Fig. \ref{fig:2_} a,d). This finding was confirmed by the Kolmogorov-Smirnov test, which demonstrated a statistically significant fit between the transformed population densities distribution and a log-normal model (Table \ref{tab:1}, Fig. \ref{fig:A}).

\begin{table}[h!]
\centering
\small
\renewcommand{\arraystretch}{1}
\caption{Statistics of original and projected population distribution}
\label{tab:1}
\begin{tabular}{|l|l|r|r|r|}
\hline
 & experiment & \multicolumn{1}{c|}{$\mu$} & \multicolumn{1}{c|}{$\sigma$} & \multicolumn{1}{c|}{p-value} \\ \hline
\multirow{2}{*}{New York City} & original & \multicolumn{1}{l|}{18.14106} & \multicolumn{1}{l|}{0.94211} & \multicolumn{1}{l|}{0.00000} \\ \cline{2-5} 
 & projected & \multicolumn{1}{l|}{18.00434} & \multicolumn{1}{l|}{1.22867} & \multicolumn{1}{l|}{\textbf{0.16439}} \\ \hline
\multirow{2}{*}{Los Angeles} & original & \multicolumn{1}{l|}{16.51779} & \multicolumn{1}{l|}{1.07319} & \multicolumn{1}{l|}{0.00000} \\ \cline{2-5} 
 & projected & \multicolumn{1}{l|}{16.58208} & \multicolumn{1}{l|}{1.12083} & \multicolumn{1}{l|}{\textbf{0.15659}} \\ \hline
\multirow{2}{*}{Chicago} & original & 16.49171 & 1.04572 & \textbf{0.38367} \\ \cline{2-5} 
 & projected & 16.36321 & 1.09286 & 0.23769 \\ \hline
\multirow{2}{*}{Houston} & original & 15.57820 & 0.80799 & 0.00027 \\ \cline{2-5} 
 & projected & 15.91225 & 0.94239 &\textbf{0.27231} \\ \hline
\multirow{2}{*}{Boston} & original & 16.18560 & 1.36385 & 0.00937 \\ \cline{2-5} 
 & projected & 15.71326 & 1.07784 & 0.29775 \\ \hline
\multirow{2}{*}{Phoenix} & original & 15.67573 & 1.06932 & 0.00000 \\ \cline{2-5} 
 & projected & 15.92882 & 1.13898 & \textbf{0.05979} \\ \hline
\multirow{2}{*}{Philadelphia} & original & 16.66367 & 0.82004 & 0.04569 \\ \cline{2-5} 
 & projected & 16.06798 & 1.08328 & \textbf{0.30635}\\ \hline
\multirow{2}{*}{San Antonio} & original & 15.33254 & 0.98614 & 0.00000 \\ \cline{2-5} 
 & projected & 15.75286 & 0.98890 & \textbf{0.90413}\\ \hline
\multirow{2}{*}{San Diego} & original & 15.77723 & 1.18909 & 0.00000 \\ \cline{2-5} 
 & projected & 15.99930 & 1.36279 & \textbf{0.26208} \\ \hline
\multirow{2}{*}{Dallas} & original & 15.53368 & 0.79847 & 0.00460 \\ \cline{2-5} 
 & projected & 15.90930 & 1.12094 & \textbf{0.88463} \\ \hline
\multirow{2}{*}{San Jose} & original & 16.11275 & 0.91071 & 0.00000 \\ \cline{2-5} 
 & projected & 15.78523 & 1.09311 & \textbf{0.81343} \\ \hline
\multirow{2}{*}{Austin} & original & 15.23158 & 1.14824 & 0.00397 \\ \cline{2-5} 
 & projected & 15.33943 & 1.13065 & \textbf{0.51603} \\ \hline
\end{tabular}
\end{table}
\newpage

Analyzing the population density map within the reshaped space, we observed a more balanced population spatial distribution compared to the original geography. In the original space (Fig. \ref{fig:2_}b), the Voronoi tessellation displays large, irregular regions with population density concentrated unevenly. This layout reflects the physical constraints and uneven development patterns inherent to the actual geographic space. However, in the reshaped embedding space (Fig. \ref{fig:2_}e), the Voronoi regions become more compact and uniformly distributed around high-commute zones. This transformation aligns population density more closely with commuting dynamics (Fig. \ref{fig:2_}c), pulling densely connected tracts together and smoothing out spatial irregularities. The commute flow network in the reshaped space \ref{fig:2_}f) also shows a more cohesive and organized structure, capturing the intrinsic connectivity of the urban layout driven by mobility patterns rather than physical distances, thus revealing a hidden, optimized urban shape. The maps for other cities can be found in Fig. \ref{fig:2A} and Fig. \ref{fig:3A}.

In addition to our primary analysis, we conducted an experimental simulation to understand urban growth patterns from a theoretical perspective. Utilizing the random-walk preferential attachment model, we simulated the settlement of a hypothetical city with a population of 100.000 residents on an unconstrained lattice. This simulation was designed to mimic the organic development of urban areas where new inhabitants settle near existing populated areas with a probability proportional to the number of nearby residents, reflecting the natural tendency for communities to cluster around established social and economic hubs.

The simulation ran iteratively, allowing the city's residential and job distributions to evolve freely without pre-set zoning restrictions or geographical barriers, thus providing a unique viewpoint on how cities might develop under purely stochastic growth conditions influenced by initial and ongoing social dynamics. After reaching the population threshold, we analyzed the resultant spatial distributions of residences and workplaces using the Kolmogorov-Smirnov test to assess their conformity to a log-normal distribution. The p-values obtained from this test were 52.08\% for residential population distribution and 13.95\% for job distribution. 

The relatively high p-value for distributions indicates a strong match with the log-normal law, implying that the emergent urban landscape, although idealized, mirrors realistic patterns where areas with higher initial settlements attract more newcomers, creating a feedback loop that enhances local growth. 

\section{Methods}\label{sec3}

In this study, we employed a random walk simulation to model the movement and settlement behaviors of individuals within a hypothetical urban grid in a toroidal space. Each point on the grid represents a potential residential or workplace location. Individuals randomly move to neighboring grid points and make decisions to settle based on the local population density, following a preferential attachment mechanism.

Initially, individuals are placed randomly across the grid. As the simulation progresses, each individual decides whether to settle at their current location or continue their random walk. The decision to settle or accept a job at a given location is probabilistically determined by assessing the likelihood of forming connections with existing residents or job opportunities, with each attachment opportunity selected according to a fixed probability parameter set as a model hyperparameter. The higher the current population or job density, the more likely it is for individuals to settle or get employed there. This decision-making process incorporates the influence of both the immediate population and the broader neighborhood, reflecting a balance between local interactions and wider urban dynamics.

The cumulative result of these individual decisions over time is the emergence of a log-normal distribution of population density across the grid. This distribution arises naturally as more populated areas attract even more residents, mimicking the 'rich-get-richer' phenomenon observed in urban settings. The simulation iterates until all individuals have settled, allowing us to observe how simple rules of movement and local interaction lead to complex population patterns characteristic of real-world cities.

\section{Conclusion}\label{sec4}

This study introduces a modeling framework for urban commute flows that enables a redefinition of urban locations by embedding them into a latent space optimized for mobility patterns. This projection unveils a "hidden" urban shape where the population distribution tends to closely follow a log-normal pattern across a wide range of U.S. cities. This uniformity suggests an underlying principle of urban organization, where population distributions, when unconstrained by geographic boundaries, naturally evolve toward log-normality. Our findings suggest that this "ideal" urban form is more frequently realized in the projected embedding space than in the original geographic layout, where historical, geographic, and regulatory factors often disrupt this statistical regularity.

The application of a random walk preferential attachment model further elucidates the conditions under which log-normal distributions emerge, revealing that population densities conform to this distribution under idealized, unconstrained settlement conditions. These results imply that historic cities, with their more varied and established structures, diverge more from this ideal shape compared to newer, planned urban areas that tend to approximate it.

Our approach has broader implications for urban planning, suggesting that latent, mobility-informed representations of cities can reveal structural regularities that physical layouts may obscure. By identifying these uniform principles, urban planners can gain insights into functional zones and population clustering patterns, potentially guiding infrastructure and service distribution decisions that align with underlying population and mobility dynamics. This framework could serve as a foundation for more resilient urban designs that harness the inherent "shape" of cities as they emerge from collective mobility patterns.

\bmhead{Acknowledgements}

This work was authored (in part) by researchers from the National Renewable Energy Laboratory (NREL), operated by Alliance for Sustainable Energy, LLC, for the U.S. Department of Energy (DOE) under Contract No. DE-AC36-08GO28308. This material is based upon work supported by the U.S. Department of Energy’s Office of Energy Efficiency and Renewable Energy (EERE) under the Vehicle Technology Program Award Number DE-EE0009211. The views expressed herein do not necessarily represent the views of the U.S. Department of Energy or the United States Government. The U.S. Government retains and the publisher, by accepting the article for publication, acknowledges that the U.S. Government retains a nonexclusive, paid-up, irrevocable, worldwide license to publish or reproduce the published form of this work, or allow others to do so, for U.S. Government purposes. 

\begin{appendices}

\section{Reshaped Cities}\label{secA1}

\begin{figure}[H]
\centering
  \includegraphics[width=1\textwidth]{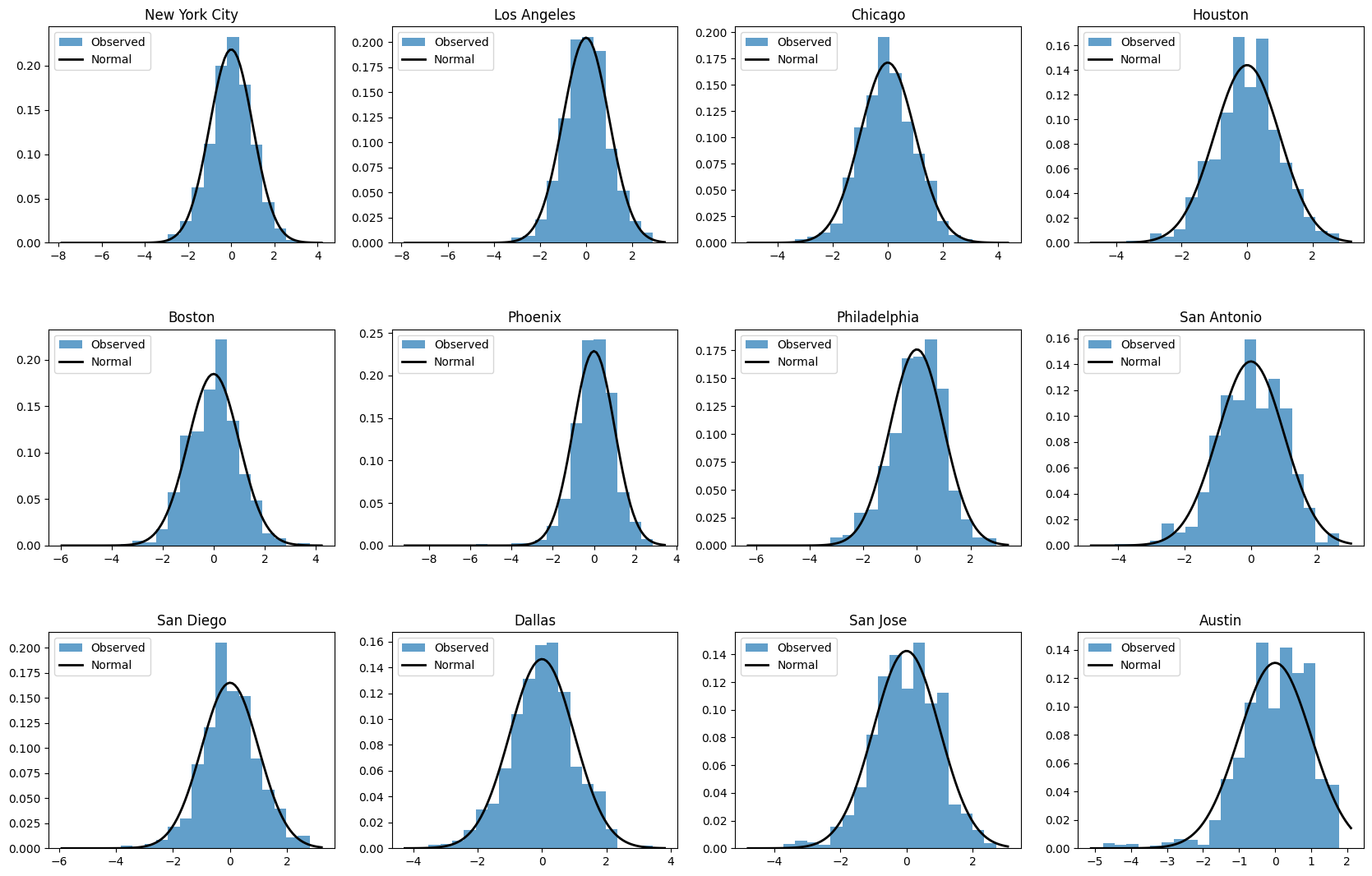}
\caption{Population log-density distribution with fitted normal curve}
\label{fig:A} 
\end{figure}

\begin{figure}[H]
\centering
  \includegraphics[width=1\textwidth]{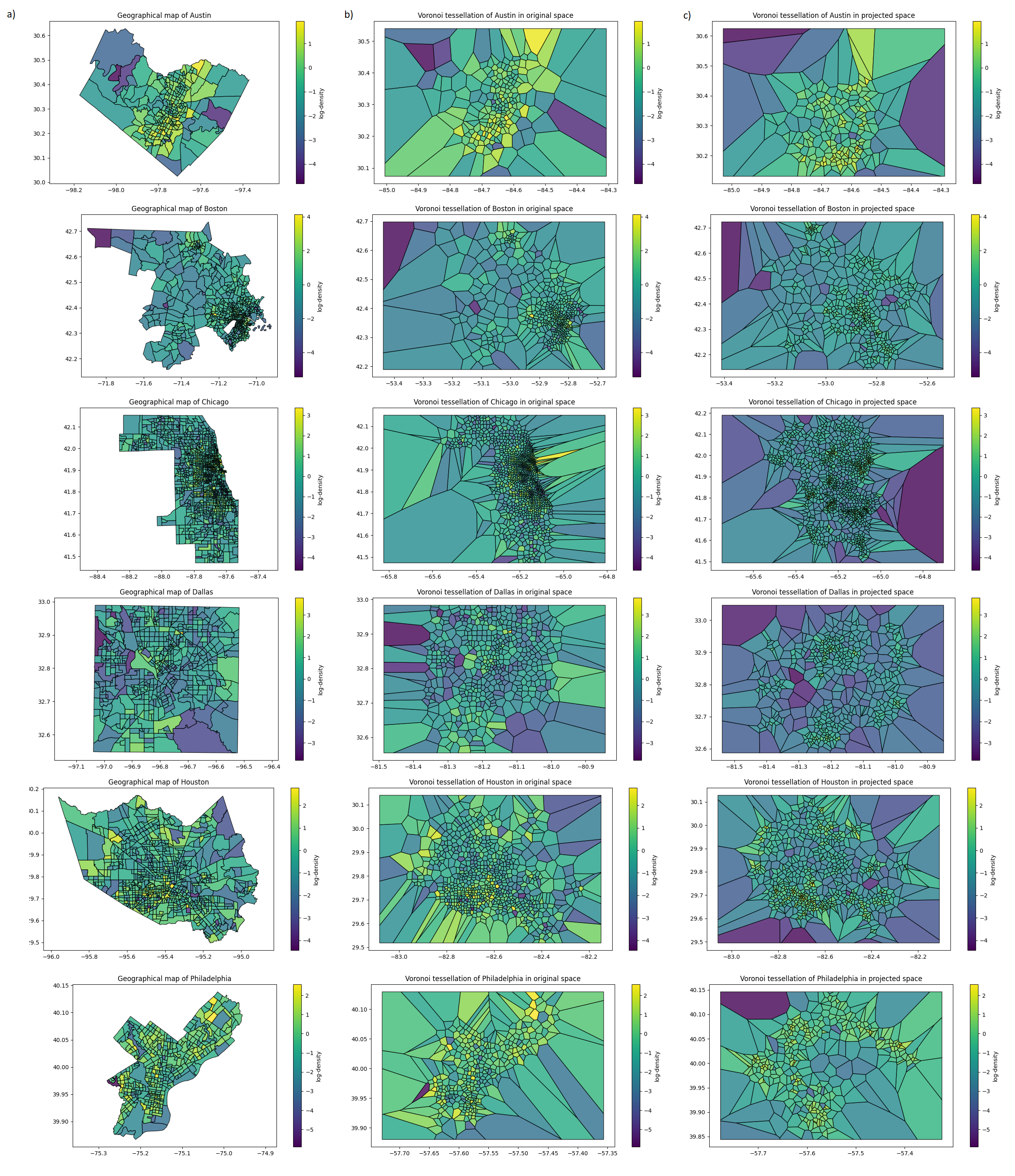}
\caption{Maps of the original cities layout (a-b), compared to their layout within the projected embedding space (c) (part 1)}
\label{fig:2A}       
\end{figure}

\begin{figure}[H]
\centering
  \includegraphics[width=1\textwidth]{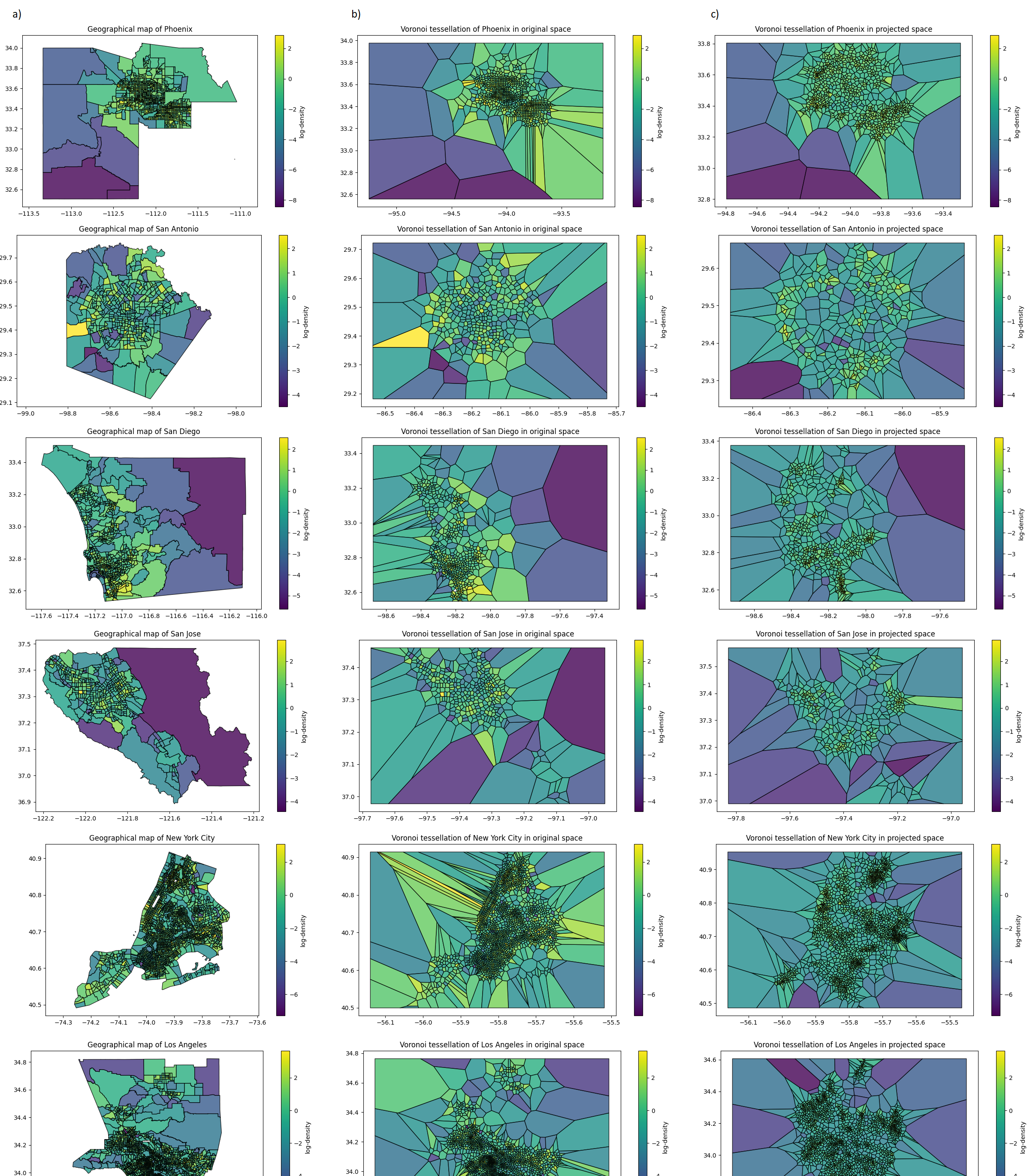}
\caption{Maps of the original cities layout (a-b), compared to their layout within the projected embedding space (c) (part 2)}
\label{fig:3A}       
\end{figure}




\end{appendices}
\newpage



\end{document}